\documentclass[]{spie}  

\usepackage{amsmath,amsfonts,amssymb}
\usepackage[colorlinks=true, allcolors=blue]{hyperref}
\usepackage{listings}
\usepackage{graphicx}
\usepackage{float}

\title{An Open-Source Gaussian Beamlet Decomposition Tool for Modeling Astronomical Telescopes}

\author[a]{Jaren N. Ashcraft}
\author[b]{Ewan S. Douglas}
\affil[a]{Wyant College of Optical Sciences, University of Arizona, Tucson, AZ 85721}
\affil[b]{Steward Observatory, University of Arizona, Tucson, AZ 85721}

\authorinfo{Further author information: (Send correspondence to J.N.A.)\\A.A.A.: E-mail: jashcraft@email.arizona.edu}

\newcommand{\A}{\bar{A}}

\begin{document} 
\maketitle

\begin{abstract}

In the pursuit of directly imaging exoplanets, the high-contrast imaging community has developed a multitude of tools to simulate the performance of coronagraphs on segmented-aperture telescopes. As the scale of the telescope increases and science cases move toward shorter wavelengths, the required physical optics propagation to optimize high-contrast imaging instruments becomes computationally prohibitive. Gaussian Beamlet Decomposition (GBD) is an alternative method of physical optics propagation that decomposes an arbitrary wavefront into paraxial rays. These rays can be propagated expeditiously using ABCD matrices, and converted into their corresponding Gaussian beamlets to accurately model physical optics phenomena without the need of diffraction integrals. The GBD technique has seen recent development and implementation in commercial software (e.g. FRED, CODE V, ASAP)\cite{Harvey15,stone_modeling_2009,koshel_novel_2001} but appears to lack an open-source platform. We present a new GBD tool developed in Python to model physical optics phenomena, with the goal of alleviating the computational burden for modeling complex apertures, many-element systems, and introducing the capacity to model misalignment errors. This study demonstrates the synergy of the geometrical and physical regimes of optics utilized by the GBD technique, and is motivated by the need for advancing open-source physical optics propagators for segmented-aperture telescope coronagraph design and analysis. This work illustrates GBD with Poisson's spot calculations and show significant runtime advantage of GBD over Fresnel propagators for many-element systems.

\end{abstract}


\section{INTRODUCTION}
 The purpose of a coronagraph is to block starlight, but allow the light from an exoplanet to pass to a detector. Many varieties of coronagraphs are employed on the ground and are planned in space, but the common denominator between every variant is that they require exceptional optical models to accurately simulate approximately a part-in-one-billion sensitivity to Earthlike planets. Coronagraphs operate principally utilizing elements that interact with the wave nature of light. Consequently, current coronagraph design tools compute diffraction integrals of large arrays to accurately simulate the near-field (Fresnel Integral, Angular Spectrum) and the far-field (Fraunhofer Integral) performance\cite{krist_2007,Soummer:07,Doug18}. 

\begin{figure}[H]
    \centering
    \includegraphics[width=\textwidth]{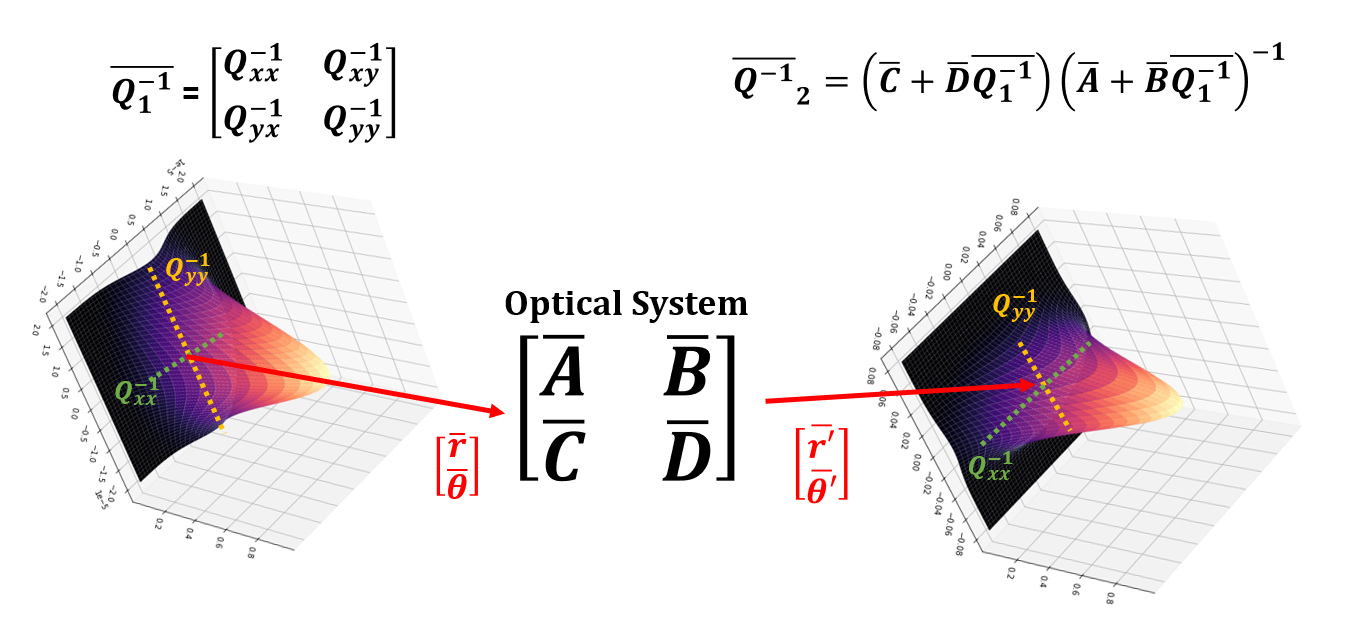}
    \caption{Gaussian beamlets are defined by a paraxial "central ray" (shown in red) that locates the Gaussian beam, and a complex curvature matrix that tracks the waist radius and curvature ($Q_{1}^{-1}$). Utilizing the linear laws of paraxial ray tracing, physical optics phenomena can be modeled.}
    \label{gbprop}
\end{figure}

A possible improvement on current simulations, Gaussian Beamlet Decomposition (GBD) is an expeditious method of propagating a wavefront via a superposition of Gaussian beams as complex rays. This enables simultaneous modeling of diffraction\cite{harvey_modeling_2015}, tilt/decenter errors\cite{Shaomin85}, and polarization\cite{worku_vectorial_2017}. An arbitrary wavefront is decomposed into a finite set of Gaussian beamlets that, when coherently added, recreate the original wavefront. The generally astigmatic beamlet is fully parameterized by a central ray that tracks the position of a beamlet, and a complex curvature matrix that describes its waist radius and curvature. Both quantities can be propagated using geometrical raytracing(Fig \ref{gbprop}), enabling fast diffraction calculations. GBD's capacity to mimic the accuracy of Fresnel diffraction was recently presented by Harvey et al \cite{Harvey15}. In this work the authors demonstrated that GBD can model both far-field (point-spread functions, interferometers) and near-field (defocused point-spread functions, spot of arago) optical phenomena. 

Gaussian beamlets are particularly useful in applications where absolute phase and diffraction effects are both of interest. For example, in very high numerical aperture systems (e.g. microscopes) polarization effects limit the accuracy of a typical scalar diffraction treatment\cite{mansuripur_distribution_1986}. GBD can simultaneously model the scalar\cite{Harvey15,Worku:18} and vector\cite{worku_vectorial_2017} nature of optical fields, resulting in a more accurate physical optics model. GBD has largely seen implementation in commercial optical design software (FRED, CODE V, ASAP) where it has been used as an analysis feature in a variety of optical science investigations. In laser applications where gaussian beams are common, GBD has been applied to study inter-cavity laser beam shaping\cite{koshel_novel_2001}. Astronomical telescope designers utilize GBD to model interferometric instrumentation\cite{Dhabal17,stone_modeling_2009} and objectives with Gaussian foci in the far-infrared/milimeter wavelengths\cite{narayananl_gaussian_nodate}. Recent efforts by Breckenridge and Harvey\cite{breckinridge_exoplanet_2018,Breck19} utilized GBD in FRED to study the diffraction features for different segmented-aperture geometries suited to optical telescopes outfitted with coronagraphs for exoplanet detection. 

In this manuscript we describe the development of an open-source GBD module to provide telescope and coronagraph designers with a new tool for modeling optical systems. The theory of GBD is introduced, and the accuracy of the model in comparison to Fresnel Diffraction is presented, with further discussion on the potential advantages of developing this platform of physical optics propagation in an open-source environment. 

\subsection{Parameters of the Gaussian Beam}
The Gaussian Beam is a solution to the Helmholtz equation that takes the form \cite{goodman17}
\begin{equation}
	V = \frac{V_o}{q(z)}exp[ik\frac{r^2}{2q(z)}]
\end{equation}

Where $V_o$ is the amplitude, $k$ is the wavenumber, $r$ is the radial coordinate in the plane perpendicular to propagation, and $q(z)$ is the complex valued constant that describes the beam's $1/e$ field size (the "waist" $w_o$) and curvature. This constant is referred to as the \textit{complex beam parameter}.
\begin{equation}
	q(z)^{-1} = \frac{1}{R(z)}+i\frac{\lambda}{\pi w(z)^2}
\end{equation}	

$q(z)$ is a convenient expression of the Gaussian beam because it fully encapsulates the information required to describe the transverse electric field of the beam as it propagates. The real part of $q(z)$ is related to the radius of curvature ($R(z)$) of the wavefront.
\begin{equation}
	R(z) = z(1+(\frac{Z_o}{z})^2)
\end{equation}
Where $Z_o$ is the rayleigh range and $z$ is the longitudinal propagation distance.
The imaginary part is related to the beam waist radius ($w(z)$)
\begin{equation}
	w(z) = w_o\sqrt{1+(\frac{z}{Z_o})^2}
\end{equation}
For a generally astigmatic beamlet with different complex curvatures in orthogonal spatial dimensions, it is convenient to define $q(z)$ as a 2x2 matrix $\overline{Q}$:

\begin{equation}
    \overline{Q}^{-1} = 
    \begin{bmatrix}
    q_{xx}^{-1} & q_{xy}^{-1} \\
    q_{yx}^{-1} & q_{yy}^{-1} \\
    \end{bmatrix}
\end{equation}

The full expression of the generally astigmatic Gaussian beam is therefore:

\begin{equation}
    V_{1} = V_{o}exp[ \frac{-ik}{2} \vec{r}^{T} \overline{Q}^{-1}\vec{r} ]
\end{equation}

\subsection{ABCD Ray Transfer Matrices}
In the regime of geometrical optics, a generally skew ray can be traced through a system using 4x4 ABCD ray transfer matrices\cite{Brouwer64}. These matrices model simple optical elements (e.g. thin lenses) with ease by operating on an input column vector that represents a light ray. The simplest ray transfer matrix that describes a paraxial and orthogonal optical system is a 2x2 operator that maps an input ($i$) spatial and angular coordinate to the appropriate output ($o$).

\begin{equation}
    \begin{pmatrix}
    y_o \\
    y'_o \\
    \end{pmatrix}
    =
    \begin{pmatrix}
    A & B \\
    C & D \\
    \end{pmatrix}
    \begin{pmatrix}
    y_i \\
    y'_i \\
    \end{pmatrix}
\end{equation}

Where $y$ is the spatial coordinate transverse to the propagation direction, and $y'$ is the angle in that dimension. In the orthogonal description, the elements of the ABCD matrix are real-valued scalars. To account for skew ray paths, the position and angle in the dimension orthogonal to $y$ and the direction of propagation must be tracked, adding two dimensions to the matrix calculus. A nonorthogonal system with tilts and decenters that map generally skew input rays to generally skew output rays is described by a 4x4 ABCD matrix. 

\begin{equation}
    \begin{pmatrix}
    x_o \\
    y_o \\
    x'_o \\
    y'_o \\
    \end{pmatrix}
    = 
    \begin{pmatrix}
    A_{xx} & A_{xy} & B_{xx} & B_{xy} \\
    A_{yx} & A_{yy} & B_{yx} & B_{yy} \\
    C_{xx} & C_{xy} & D_{xx} & D_{xy} \\
    C_{yx} & C_{yy} & D_{yx} & D_{yy} \\
    \end{pmatrix}
    \begin{pmatrix}
    x_i \\
    y_i \\
    x'_i \\
    y'_i \\
    \end{pmatrix}
\end{equation}

For simplicity, it is convenient to represent that radial position in the plane transverse to propagation (x,y) and the corresponding angle in the dimension (x',y') as a position and angle vector respectively ($\vec{r},\vec{\theta}$). The ABCD matrix can similarly be condensed into 2x2 sub matrices that operate on each spatial dimension, yielding a familiar notation.

\begin{equation}
    \begin{pmatrix}
    \vec{r_{o}} \\
    \vec{\theta_{o}} \\
    \end{pmatrix}
    =
    \begin{pmatrix}
    \overline{A} & \overline{B} \\
    \overline{C} & \overline{D} \\
    \end{pmatrix}
    \begin{pmatrix}
    \vec{r_{i}} \\
    \vec{\theta_{i}} \\
    \end{pmatrix}
\end{equation}

This description is powerful because it communicates the elegance and simplicity of ray transfer matrices. All dimensions transverse to propagation are accounted for, but the calculus to propagate a ray is still the same. Similarly, the complex curvature of a generally astigmatic Gaussian beamlet can be propagated through a nonorthogonal optical system using the 4x4 ABCD ray transfer matrix

\begin{equation}\label{qmult}
    \overline{Q_{2}}^{-1} = (\overline{C} + \overline{D}\overline{Q_{1}}^{-1})(\overline{A} + \overline{B}\overline{Q_{1}}^{-1})^{-1}
\end{equation}

This property of the complex beam parameter matrix is the cornerstone of Gaussian beamlet propagation. The waist radius, beam curvature, and position can all be propagated using the linear laws of geometrical ray tracing. This investigation will consider the propagation of Gaussian beamlets through thin lens elements ($\overline{O}$) and distances in a homogenous index of refraction ($\overline{D}$), but other components (e.g. GRIN media, tilted/decentered elements) have ABCD matrix equivalents that can be easily incorporated\cite{Shaomin85}. 

\begin{equation}
    \overline{O} = 
    \begin{pmatrix}
    1 & 0 & 0 & 0 \\
    0 & 1 & 0 & 0 \\
    1/efl_x & 0 & 1 & 0 \\
    0 & 1/efl_y & 0 & 1 \\
    \end{pmatrix}
    ;                
    \overline{D} = 
    \begin{pmatrix}
    1 & 0 & d_x/n & 0 \\
    0 & 1 & 0 & d_y/n \\
    0 & 0 & 1 & 0 \\
    0 & 0 & 0 & 1 \\
    \end{pmatrix}
\end{equation}
\subsection{Complex Ray Tracing}
A powerful property of the Gaussian beam is its ability to be propagated using the linear laws of geometrical optics. 
Propagation of Gaussian beams via ray tracing was originally published by Arnaud in 1985 for laser propagation\cite{Arnaud85}.
This technique enables complex field propagation through simple matrix multiplication, rather than relying on computationally intensive Fourier transforms. The position of a Gaussian beam is tracked by a \emph{central ray} ($\vec{r_{i}}$ $\vec{\theta_{i}}$)$^{T}$ that emanates perpendicular to the beam waist center and through the peak of the beam. Propagating this ray through an ABCD Optical System matrix will return the position of the propagated beamlet. The beam waist size and radius of curvature are then propagated through the same ABCD system via the complex curvature matrix $\overline{Q_{2}}^{-1}$ using equation \ref{qmult} .

For a more general description of beamlet propagation from the input location $\vec{r_{i}}$ to the output location $\vec{r_{o}}$, we look to the rigorous mathematics of Cai and Lin\cite{cai_decentered_nodate}, who proposed the generally decentered and elliptical Gaussian Beamlet ($V_{DEGB}$). 

\begin{equation}
    V_{DEGB} = [det(\overline{A} + \overline{B}\overline{Q_{1}}^{-1})]^{1/2} exp[-ikz_{o}]*exp[-\frac{ik}{2}\vec{r_{o}}^{T}\overline{Q_{2}}^{-1}\vec{r_{o}}]*exp[\overline{X}]*exp[\overline{Y}]
\end{equation}

Where $\overline{X}$ and $\overline{Y}$ are the phase factors that arise from the beamlet decenter, reproduced below.

\begin{equation}
    \overline{X} = -\frac{ik}{2}\vec{r_{i}}^{T}(\overline{Q_{2}}^{-1} + \overline{A}^{-1}\overline{B})\vec{r_{i}}
\end{equation}

\begin{equation}
    \overline{Y} = ik\vec{r_{i}}^{T}(\overline{A}\overline{Q_{2}}^{-1} + \overline{B})\vec{r_{o}}
\end{equation}

While the derivation of this expression was described by the authors as "tedious but straightforward", it results in a powerful description of a key element in the GBD method: The ability to spatially decompose an incident wavefront.

\subsection{Gaussian Beamlet Decomposition}
    
    The decomposition of a wavefront into Gaussian beamlets requires a treatment of how the phase evolves with the propagation of a decentered beamlet. This is because typical decomposition algorithms accomplish wavefront decomposition through a distribution of beamlets in the entrance pupil of the optical system. Few methods have been explored to best decompose an input wavefront into a finite set of Gaussian beams. The simplest method is an even distribution of beams across the input wavefront (henceforth referred to as \emph{even sampling}). The number of Gaussian beamlets ($N_{gb}$) across one dimension ($W$) of the wavefront is given by
    
    \begin{equation}
        N_{gb} = \frac{W*OF}{2w_{o}}.        
    \end{equation}
    
    Where $OF$ is the overlap factor of the beamlets. This factor describes the degree to which adjacent beamlets overlap. Some partial overlap of adjacent beams is desirable to reduce the ripple that results from undersampling the wavefront. However, an $OF$ that is too high will result in a soft aperture edge, which low-pass filters the input wavefront. Literature suggests that an $OF=1.5-1.7$ \cite{Harvey15} is the appropriate range to experience the ripple and low-pass filtering minimally. An alternative sampling method which yields a more efficient decomposition for optical systems with rotational symmetry evenly distributes the beamlets on a Fibonacci spiral (henceforth referred to as \emph{Fibonacci sampling}). 
    
    \begin{figure}[H]
        \centering
        \includegraphics[width=0.65\textwidth]{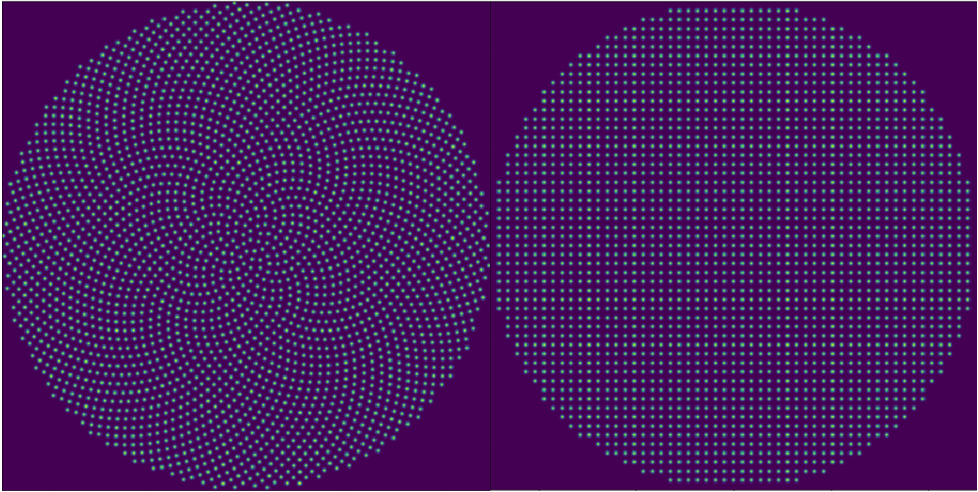}
        \includegraphics[width=0.65\textwidth]{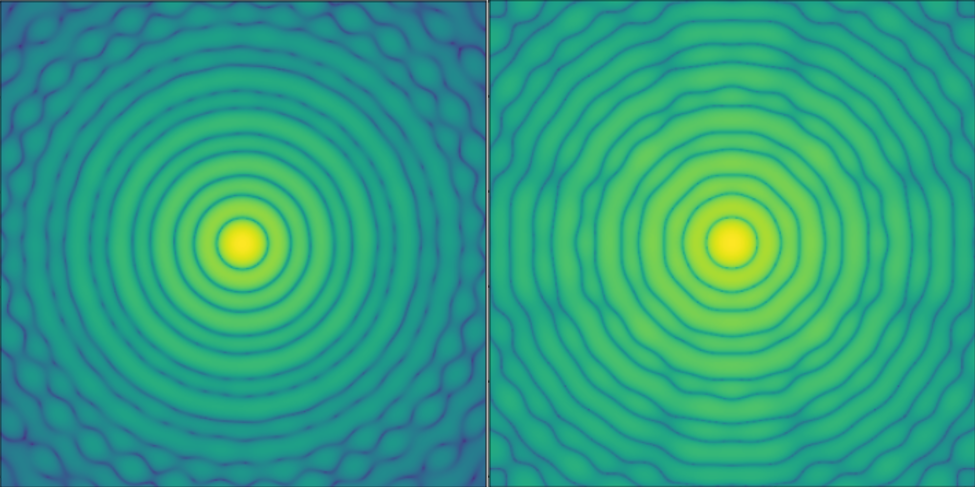}
        \caption{Demonstration of various sampling schemes (Right) Even sampling\cite{Harvey15} (Left) Fibonacci sampling\cite{Worku:18}. (Top) Undersampling in pupil plane to show beamlet distribution, (Bottom) The corresponding PSF.}
        \label{fig:my_label}
    \end{figure}
    
    Both sampling schemes have been developed in the GBD code for comparison, given that each may be more optimal for different system geometries. The Fibonacci sampling scheme is interesting because it has been shown to result in lower RMS error to a flat wavefront for circular entrance pupils, which are very common for optical systems \cite{Worku17}.
    
\newpage
\section{METHODS}
    In order to validate our approach, we baseline our systems in POPPY (Physical Optics Propagation in Python). POPPY is an open-source project that simulates the behavior of complex scalar optical fields using numerical FFT-based approaches \cite{Perrin12}. This project was originally developed as the physical optics engine for WebbPSF, a tool developed to simulate the point spread functions of the James Webb Space Telescope (JWST) using Fraunhofer diffraction. The capacity of this tool was later extended\cite{Doug18} to incorporate Fresnel diffraction\cite{lawrence_optical_1992}, which captures both the near and far-field effects that limit high-contrast imaging systems\cite{krist_2007}. This implementation was used to model the Magellan Extreme Adaptive Optics upgrade (MagAO-X\cite{lumbres_modeling_2018}) but in order to model all the surfaces in the system the run time is significant ($\gtrapprox$ 1 min per run).
    
    An object-oriented approach to Gaussian beamlet decomposition was written in Python similarly to POPPY's Fresnel mode for demonstration. A Gaussian beamlet Wavefront class and an Optical System class were added, with methods included to sufficiently generate an orthogonal optical system composed of paraxial lenses with vignetting apertures.
    
    \begin{table}[H]
        \centering
        \begin{tabular}{|c|c|}
        \hline
            OpticalSystem Method & Description  \\
            \hline
            add$\_$optic(efl) & Adds a thin lens to ABCD system matrix \\
            add$\_$distance(distance,index) & Adds a propagation distance d/n \\
            add$\_$aperture(shape,size) & Add a ray-vignetting aperture \\
            add$\_$detector(size,npix) & Adds an analysis plane where diffraction calculation is done \\
            \hline
        \end{tabular}
        \caption{Description of current optical system methods for the GBD code.}
        \label{tab:my_label}
    \end{table}
    
    \begin{lstlisting}[language=python]
    # Initialize the Gaussian beamlet wavefront and optical system
    gwfr = GaubletWavefront(wavelength=2.2e-6,size=2.4,samplescheme='fibonacci')
    osys = GaubletOpticalSystem(epd=2.4,dimd=5e-4,npix=512)
    
    # Add a circular optic of 5.52 meter focal length
    osys.add_aperture(shape='lyot',diameter=2.4)
    osys.add_optic(efl = 5.52)
    
    # Propagate to focus and add a detector
    osys.add_distance(distance=5.52,index=1)
    osys.add_detector()
    \end{lstlisting}
    
    \begin{figure}[H]
        \centering
        \includegraphics[width=0.74\textwidth]{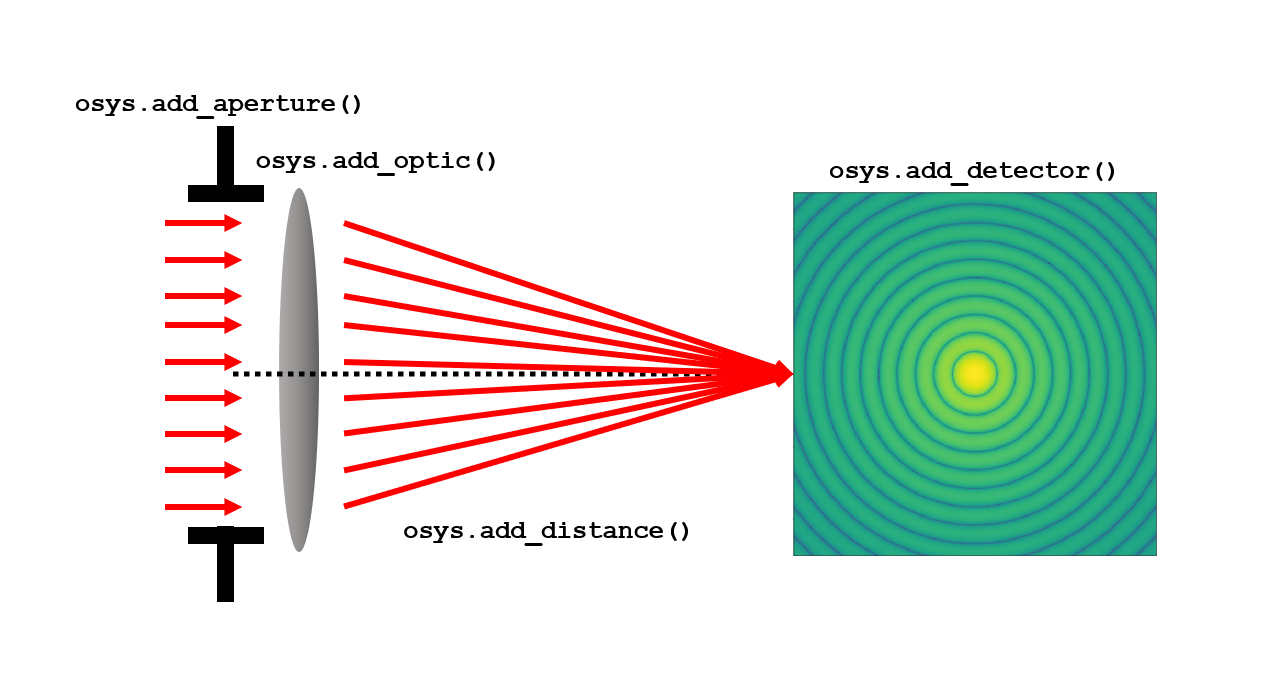}
        \caption{(Top) Sample code to demonstrate the user-friendly \& object oriented construction of the GBD module and (Bottom) a graphical representation of what the code creates.}
        \label{fig:my_label}
    \end{figure}

	\subsection{Accelerated Computing}
	
	Utilizing the iPython \emph{timeit} function the propagator was profiled to identify the components that were critically slowing runtime. The GBD code relies on computing the exponential of an array with dimensions [$\#$ pixels,$\#$ pixels,$\#$ beamlets], as well as a for loop to construct the phase of each beamlet. For a large number of beamlets, both numpy's exponential and the phase construction loop become computationally prohibitive. To mitigate the effects on runtime, two alternative acceleration methods were investigated. Numexpr\cite{cooke_david_2018_1492916} is an accelerated mathematical computing package for Python tailored to the processing of large arrays. Rather than handling large arrays all at once numexpr computes the array in chunks to optimize computing time. Multi-threading is also viable in numexpr for further enhancing performance. Numba\cite{lam_numba_2015} is another accelerator for large array computation that achieves performance enhancement by translating a numba decorated function into machine code for increased computing performance. Numexpr produced the more expeditious exponential function (1.65x Numba's time), but is not able to process for loops. The phase construction loop was written as a function that numba can accelerate, resulting in minimized computation time. By taking advantage of multi-threading computation can be accelerated further. For 90 threads the exponential of a complex array (dimension [512,512,2000]) was 1000x faster in numexpr than with numpy's exponential on the same machine.

\section{RESULTS}

    \subsection{Far Field - The Airy Disk}
    
    One of the first problems introduced to students of physical optics is the solution to the Fraunhofer diffraction integral for a circular aperture. This yields the familiar airy pattern, which is the basis for all images produced by rotationally symmetric systems. Consequently, it is a powerful example for the comparison of different approaches to diffraction calculations. The Fresnel and Fraunhofer diffraction integrals both have valid solutions to the focal plane intensity. For sufficient validation of GBD's far-field performance, we present a comparison of the airy pattern generated by an ideal lens operating at a focal ratio of $F/2.3$ at a wavelength of $\lambda = 2.2\mu m$. 
    
    \begin{figure}[H]
        \centering
        \includegraphics[width=0.7\textwidth]{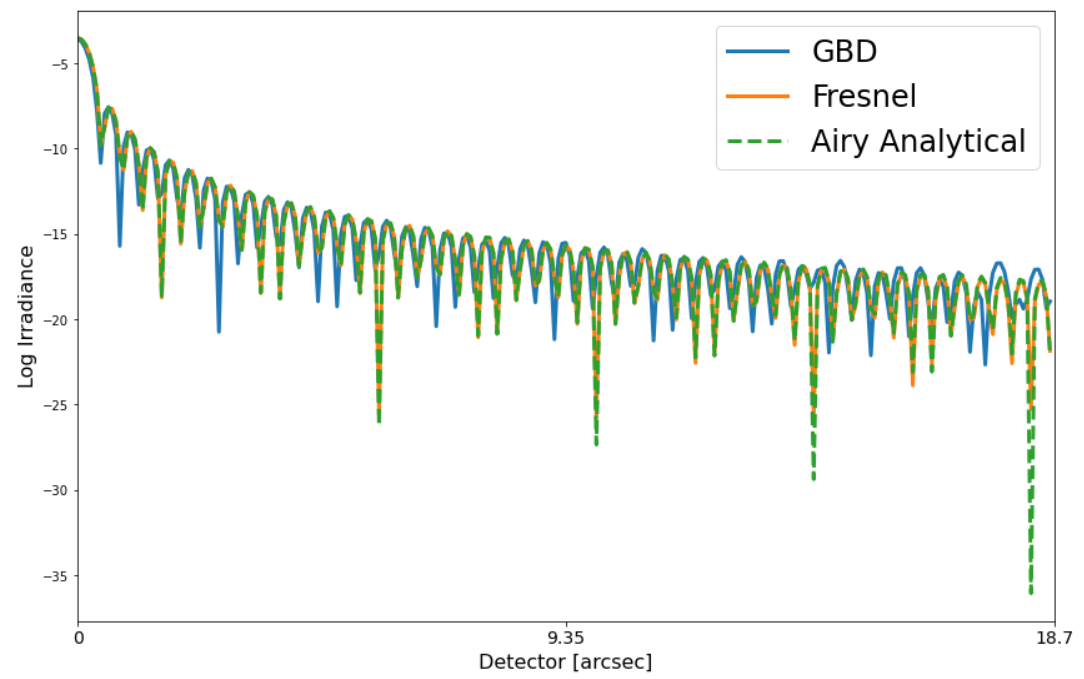}
        \caption{Cross-section of the focal plane intensity on a log scale generated by an ideal F/2.3 lens using Gaussian Beamlets (blue) and Fresnel (orange) in comparison to the analytical airy function (green).}
        \label{psflog}
    \end{figure}
    
    The similarities in PSF structure between the three regimes of diffraction out to nearly the 50th airy ring are compelling and indicative of GBD's ability to preserve high spatial frequency content. This cements GBD as a valid physical optics method for far-field diffraction calculations.
    
    \subsection{Near Field - Defocused PSF}
	The near-field regime of diffraction is much more challenging to model, but can capture many interesting physical optics phenomena. One interesting feature is the behavior of the on-axis intensity of an airy disk as defocus is added. Defocus ($W_{020}$) is a field-independent aberration that scales quadratically with the pupil size. The distance ($L$) corresponding to $W_{020}$ waves of defocus is:
	
	\begin{equation}
	    L = 8 W_{020}\lambda F^2
	\end{equation}
	Where F is the F-number of the optic.
	When light travels integer ($n$) waves of $W_{020}$ from focus, the on-axis irradiance goes to zero. Conversely, when light travels half-integer ($(n+1)/2$) waves of $W_{020}$ from focus, the on-axis irradiance maximizes. This can typically only be modeled by direct propagation using Fresnel or Angular Spectrum diffraction integrals. However, GBD is also capable of producing the same results.
	
    	\begin{figure}[H]
    	    \centering
    	    \includegraphics[width=\textwidth]{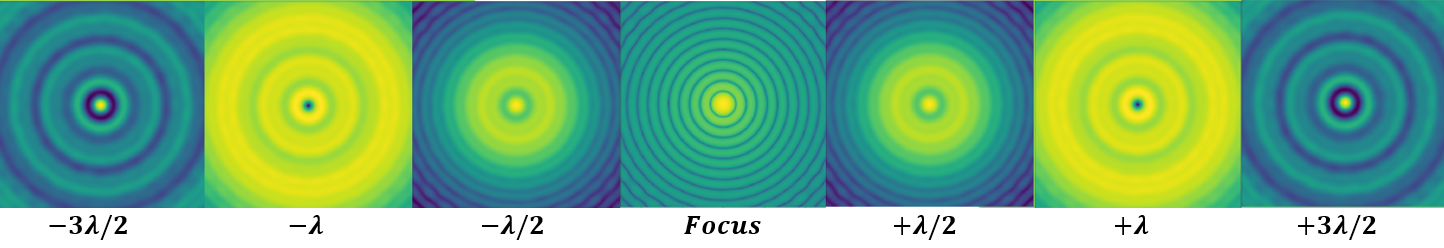}
    	    \caption{Field intensity distribution near the focal plane at $n\lambda/2$ waves of defocus. The on-axis irradiance is expected to maximize at half-integer waves of defocus and minimize at integer waves of defocus. Recovering this result using GBD is proof that the method is capable of accurately modeling intensity distributions near the focal plane.}
    	    \label{fig:my_label}
    	\end{figure}

	\subsection{Near Field - Poisson's Spot}
	Poisson's spot is a phenomenon that can only arise with a physical optics treatment, making it another interesting figure of comparison between existing physical optics propagators and GBD. When collimated light passes through an annular aperture, a bright spot is seen on the axis that crosses through the innermost circular obscuration. With a purely geometrical treatment this phenomenon is impossible because all light rays are parallel to the optical axis. However, by virtue of diffraction the light near the aperture edges will constructively interfere on-axis yielding the observed bright spot (Poisson's Spot)\cite{hecht_optics_2012}. 
	
    	\begin{figure}[H]
    	    \centering
    	    \includegraphics[width=\textwidth]{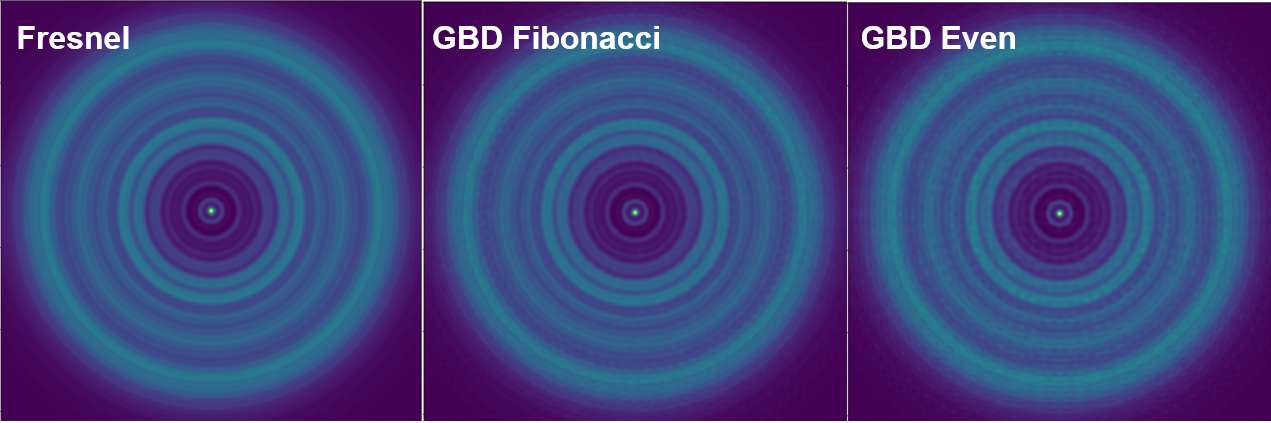}
    	    \caption{Comparison of Poisson's spot for cases of equal Fresnel number. (Left) POPPY's Fresnel propagator. (Middle) GBD propagator using Fibonacci sampling. (Right) GBD propagator using even sampling.}
    	    \label{Poisson1}
    	\end{figure}
    	
    	\begin{figure}[H]
    	    \centering
    	    \includegraphics[width=0.4\textwidth]{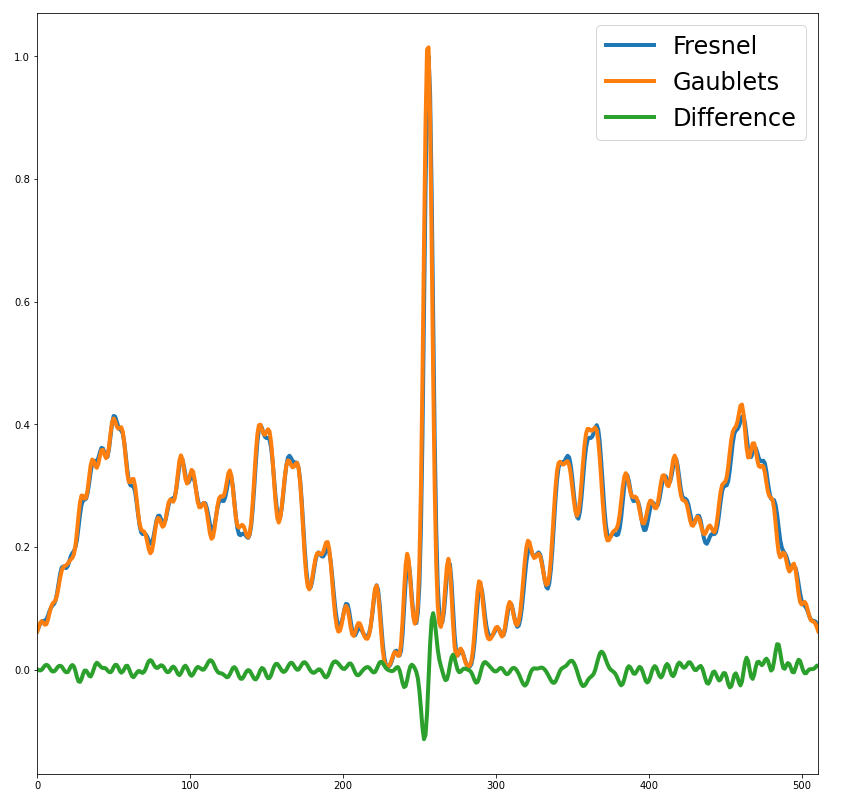}
    	    \caption{Cross-section of the Fresnel and GBD Fibonacci results from Fig \ref{Poisson1} superimposed with the difference of the Fresnel distribution with the GBD distribution. }
    	    \label{Poisson2}
    	\end{figure}
    	
    Presented above are figures (Fig. \ref{Poisson1},\ref{Poisson2}) demonstrating the accuracy of GBD's capacity to model near-field propagation and Poisson's spot relative to POPPY's Fresnel propagator. Figure \ref{Poisson1} provides a visual comparison of the sampling schemes presented in section 1.4 to illustrate the ramifications of Gausian beamlet sampling. The Fibonacci sampling scheme is clearly more suited to this problem than the even sampling scheme due to the rotational symmetry of the aperture. The even sampling leaves much more apparent features of beamlet distribution in the propagated field.
    
	\subsection{Runtime Comparison v.s. Poppy Fresnel Mode}
	
	To examine the current computational efficiency of GBD a comparison between the runtimes of POPPY's Fresnel mode and GBD at different beamlet samplings were examined. Due to the reliance on geometric propagation, GBD's runtime was fairly consistent for optical systems with 10-50 optical elements. Fresnel propagation relies on a diffraction integral to propagate to each element, so the runtime scaled up as optical elements increased.
	
	\begin{figure}[H]
	    \centering
	    \includegraphics[width=0.4\textwidth]{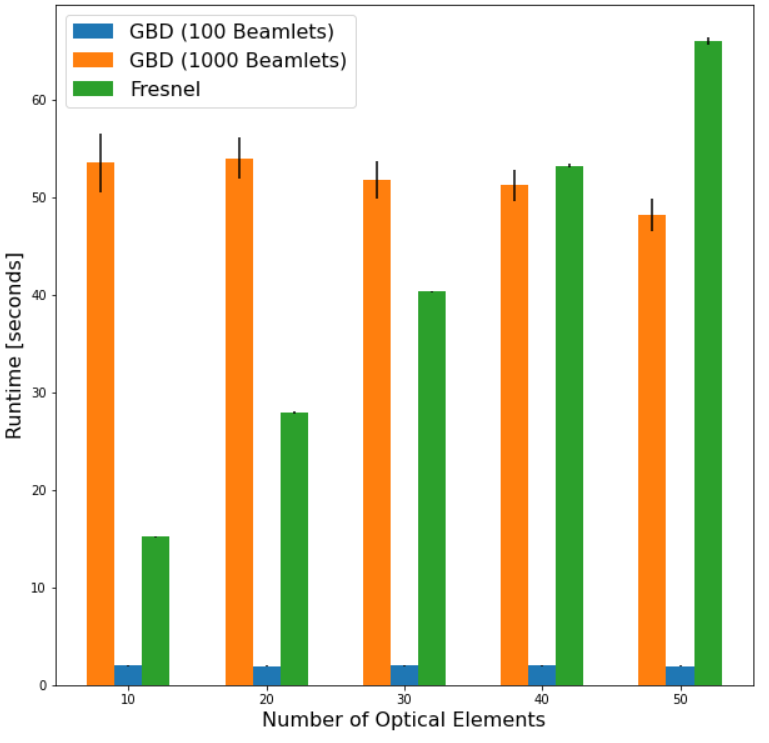}
	   \caption{Comparison of runtime between GBD  POPPY's Fresnel mode for increasing number of optical elements. Due to GBD's reliance on raytracing for propagation, the runtime is dominated by the number of beamlets used rather than the number of elements, making it ideal for many-element systems.}
	    \label{fig:runtime}
	\end{figure}

\newpage

\section{FUTURE WORK}
This work is intended to communicate the efficacy of a new physical optics propagator for open-source implementation geared toward the development of optical instrumentation for astronomical telescopes. We see potential for substantive improvement of this code in the regimes of acceleration and more comprehensive modeling of optical systems.

\subsection{Tilt/Decentered ABCD Ray Transfer Matrices}

Tilts and decenters of nonorthogonal optical elements can be accounted for by raising the rank of the ABCD matrix to 5 and adding displacement terms in position ($\delta x$,$\delta y$) and angle ($\delta x'$,$\delta y'$). To account for the increase in rank a unit valued element is added to the ray vector.

\begin{equation}
    \begin{pmatrix}
    x_o \\
    y_o \\
    x'_o \\
    y'_o \\
    1 \\
    \end{pmatrix}
    = 
    \begin{pmatrix}
    A_{xx} & A_{xy} & B_{xx} & B_{xy} & \delta x \\
    A_{yx} & A_{yy} & B_{yx} & B_{yy} & \delta y\\
    C_{xx} & C_{xy} & D_{xx} & D_{xy} & \delta x'\\
    C_{yx} & C_{yy} & D_{yx} & D_{yy} & \delta y'\\
    0 & 0 & 0 & 0 & 1 \\
    \end{pmatrix}
    \begin{pmatrix}
    x_i \\
    y_i \\
    x'_i \\
    y'_i \\
    1 \\
    \end{pmatrix}
\end{equation}

This allows for the direct modeling of tilt and decenter errors in any component of a given ABCD system matrix. Rudimentary results of a simple tilt and decenter of a thin lens have already been demonstrated by the presented GBD code, but leveraging this feature of ray transfer matrices to do tolerance-informed optimization of a physical optics model would be a powerful tool for diffraction-limited instrument design and development.

\subsection{Truncated Gaussian Beamlet Decomposition}

Accurate diffraction simulation of high spatial frequency physical optics phenomena (e.g. Fig \ref{Poisson1}) require a large number of beamlets to produce a sufficient result. This is in part due to the low-pass filtering problem of GBD. The beamlets at the edge of apertures do not fully capture the hard edge of an aperture, limiting the range of spatial frequencies that can be captured by GBD. This is also observed in the PSF comparison (Fig. \ref{psflog}) where we see a larger departure near the edge of the image (where the high spatial frequency content of the pupil is stored). Sampling the aperture with smaller beamlets mitigates the roll-off at the edges, but results in a considerable computational burdern (Fig. \ref{fig:runtime}). Worku and Gross's solution to this problem was to employ truncated Gaussian beamlets in the decomposition phase. This expression of the Gaussian beamlet can be propagated geometrically like the typical Gaussian beamlet, but captures more high spatial frequency information.

	\begin{figure}[H]
	    \centering
	    \includegraphics[width=0.9\textwidth]{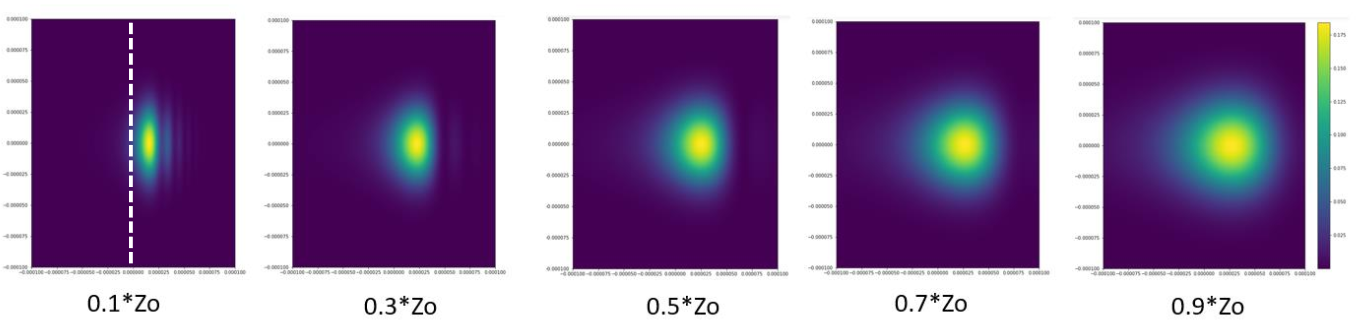}
	    \caption{Propagation of a half-truncated Gaussian beam at fractions of the Rayleigh range. Observed here are the near-field diffraction effects that result from capturing higher spatial frequency information at a hard aperture (white dashed line).}
	    \label{fig:my_label}
	\end{figure}
	
\subsection{GPU Acceleration}
Greynolds\cite{greynolds_ten_2020} recently published a manuscript indicating that there has been no implementation of GBD that leverages the computational advantages of GPU-acceleration. The presented GBD code utilizes Numba\cite{lam_numba_2015} to assemble the phase of each beamlet one at a time. However, this calculation can be done in paralell on a GPU by using NVIDIA's CUDA\cite{cuda}, which is compatible with Numba. We anticipate substantive runtime decreases that will enable higher sampling for a more comprehensive physical optics model of diffraction-limited systems.

\subsection{Alternative Sample Schemes}
Current GBD methods utilize some uniform distribution (across an aperture or Fibonacci spiral) of beamlets in the entrance pupil of the optical system. While simple to set up, this isn't necessarily the most computationally efficient sampling scheme. Xia et al\cite{xia_fast_nodate} configured an optimization-based approach to compute a best-fit generally astigmatic beamlet to the wavefront, subtract it, and repeat to generate a set of beamlets to propagate that minimize the residual error. Greynolds\cite{Greynolds14} discussed the expansion of GBD using a complete set of Hermite-Gaussian or Laguerre-Gaussian polynomials to more completely reconstruct an arbitrary wavefront in the entrance pupil. Both options are compelling for increasing the accuracy and decreasing the runtime of GBD, and will be considered for addition with the development of the GBD module.

\subsection{Polarization Ray Tracing for Vector Field Propagation}
Previous investigations into GBD\cite{Worku17} have illustrated the convenience of merging GBD's scalar diffraction capabilities with polarization ray tracing by assigning a Jones vector to the central ray to track the direction of the electric field vector. The Jones vector is expressed in three dimensions: one in the direction of the central ray, and the two directions transverse to the propagation vector (referred to as the \textbf{s,p,k} basis \cite{Yun11}). Different electric field vectors can be assigned to every beamlet used in the decomposition. This enables users to assign a spatially-varying polarization state across the input optical field, a feature not yet implemented in any open-source field propagator. This is of particular interest to the high-contrast imaging community due to the limitations polarization aberrations place on coronagraphs\cite{Breck19}. 

\subsection{POPPY Integration}
This study was motivated by the desire to expand the capacity of POPPY by adding an alternative physical optics propagation method to POPPY's suite of Fraunhofer and Fresnel modes. With further optimization, GBD's performance and runtime may approach that of POPPY's Fresnel mode. GBD also introduces the capacity to model vector field propagation and misalignment errors; features which are not present in POPPY currently. Successful formalization of the GBD code for integration into POPPY would result in a tool capable of highly comprehensive physical optics modeling.

\section{CONCLUSION}

This manuscript presents the development of a new open-source platform for Gaussian beamlet decomposition written in Python and verified against POPPY, an existing open-source physical optics code developed in the context of astronomical telescopes. We demonstrate that GBD is capable of modeling both near and far-field diffraction effects, and illustrate its potential for further development to model vector field propagation and misalignment errors through polarization ray tracing and 5x5 ABCD matrices respectively. GBD's synthesis of geometrical propagation and high-fidelity diffraction simulation make it a worthy technique for the design and analysis of future high-contrast imaging instruments\cite{douglas_review_nodate},\cite{maier_design_nodate}. A GitHub repository with the code used in this investigation is published at Jashcraf/Gaussian-Beamlets\cite{jaren_n_ashcraft_2020_4299354}. 

    

\section{Acknowledgements}
Portions of this work were supported by the Arizona Board of Regents Technology Research Initiative Fund (TRIF). An allocation of computer time from the UA Research Computing High Performance Computing (HPC) at the University of Arizona is gratefully acknowledged. The author also thanks Weslin Pullen for helpful discussion throughout the duration of this study.


\newpage
\bibliography{report,diffraction} 
\bibliographystyle{spiebib} 
\end{document}